\documentclass[aps,prb,reprint,preprintnumbers,superscriptaddress,amsmath,amssymb,bibnotes,longbibliography]{revtex4-2}

\usepackage{graphicx}
\usepackage{dcolumn}
\usepackage{epstopdf}
\usepackage{bm}
\usepackage{flushend}
\usepackage[pdfstartview=FitH, CJKbookmarks=true, bookmarksnumbered=true, bookmarksopen=true, colorlinks, pdfborder=001, linkcolor=blue, anchorcolor=blue, citecolor=blue,urlcolor=blue,breaklinks]{hyperref}
\usepackage[mathlines]{lineno}
\begin{document}
\title{Nodeless superconductivity in the charge density wave superconductor LaPt$_2$Si$_2$}
\author{Z. Y. Nie}
\affiliation{Center for Correlated Matter and Department of Physics, Zhejiang University, Hangzhou 310058, China}
\affiliation  {Zhejiang Province Key Laboratory of Quantum Technology and Device, Department of Physics, Zhejiang University, Hangzhou 310058, China}
\author{L. C. Yin}
\affiliation{Center for Correlated Matter and Department of Physics, Zhejiang University, Hangzhou 310058, China}
\affiliation  {Zhejiang Province Key Laboratory of Quantum Technology and Device, Department of Physics, Zhejiang University, Hangzhou 310058, China}
\author{A.Thamizhavel}
\affiliation{DCMP ${\emph{\&}}$ MS, Tata Institute of Fundamental Research, Mumbai 400005, India}
\author{A. Wang}
\affiliation{Center for Correlated Matter and Department of Physics, Zhejiang University, Hangzhou 310058, China}
\affiliation  {Zhejiang Province Key Laboratory of Quantum Technology and Device, Department of Physics, Zhejiang University, Hangzhou 310058, China}
\author{B. Shen}
\affiliation{Center for Correlated Matter and Department of Physics, Zhejiang University, Hangzhou 310058, China}
\affiliation  {Zhejiang Province Key Laboratory of Quantum Technology and Device, Department of Physics, Zhejiang University, Hangzhou 310058, China}
\author{L. Q. Che}
\affiliation{Center for Correlated Matter and Department of Physics, Zhejiang University, Hangzhou 310058, China}
\affiliation  {Zhejiang Province Key Laboratory of Quantum Technology and Device, Department of Physics, Zhejiang University, Hangzhou 310058, China}
\author{F. Du}
\affiliation{Center for Correlated Matter and Department of Physics, Zhejiang University, Hangzhou 310058, China}
\affiliation  {Zhejiang Province Key Laboratory of Quantum Technology and Device, Department of Physics, Zhejiang University, Hangzhou 310058, China}
\author{Z. Hossain}
\affiliation{Department of Physics, Indian Institute of Technology, Kanpur 208016, India}
\author{M. Smidman}
\affiliation{Center for Correlated Matter and Department of Physics, Zhejiang University, Hangzhou 310058, China}
\affiliation  {Zhejiang Province Key Laboratory of Quantum Technology and Device, Department of Physics, Zhejiang University, Hangzhou 310058, China}
\author{X. Lu}
\email[Corresponding author. ]{xinluphy@zju.edu.cn}
\affiliation  {Center for Correlated Matter and Department of Physics, Zhejiang University, Hangzhou 310058, China}
\affiliation  {Zhejiang Province Key Laboratory of Quantum Technology and Device, Department of Physics, Zhejiang University, Hangzhou 310058, China}
\affiliation  {Collaborative Innovation Center of Advanced Microstructures, Nanjing 210093, China}
\author{H. Q. Yuan}
\email[Corresponding author. ]{hqyuan@zju.edu.cn}
\affiliation  {Center for Correlated Matter and Department of Physics, Zhejiang University, Hangzhou 310058, China}
\affiliation  {Zhejiang Province Key Laboratory of Quantum Technology and Device, Department of Physics, Zhejiang University, Hangzhou 310058, China}
\affiliation{Collaborative Innovation Center of Advanced Microstructures, Nanjing 210093, China}
\affiliation  {State Key Laboratory of Silicon Materials, Zhejiang University, Hangzhou 310058, China}
\date{\today}

\begin{abstract}
We have studied the superconducting gap structure of LaPt$_2$Si$_2$ by measuring the temperature dependence of the London penetration depth shift $\Delta\lambda(T)$ and point contact spectroscopy of single crystals. $\Delta\lambda(T)$ shows an exponential temperature dependence at low temperatures, and the derived normalized superfluid density $\rho_{s}(T)$ can be well described by a single-gap \emph{s}-wave model. The point-contact conductance spectra can also be well fitted by an \emph{s}-wave Blonder-Tinkham-Klapwijk model, where the gap value shows a typical BCS temperature and magnetic field dependence consistent with type-II superconductivity. These results suggest fully gapped superconductivity in LaPt$_2$Si$_2$, with moderately strong electron-phonon coupling.

\begin{description}
\item[PACS number(s)]

\end{description}
\end{abstract}

\maketitle
\section{INTRODUCTION}
The interplay between charge density wave (CDW) order and superconductivity has been a frontier subject in condensed-matter physics \cite{A15,CDW2,CDWandSC1}. In a number of CDW materials \cite{CDWcompete}, superconductivity competes with the CDW order, which leads to a sharp enhancement of the superconducting transition temperature $T_c$ upon suppressing the CDW transition. Recently, there has been a growing interest focusing on the nature of the superconducting state near a possible CDW or structural quantum critical point, when a superconducting dome is observed close to the disappearance of CDW order upon applying pressure or chemical doping \cite{TaS,CuTiSe2,lpi1,CaRhSn}. This is somewhat similar to the antiferromagnetic quantum criticality in heavy fermion compounds \cite{CeIn3,CCS} and FeAs-based superconductors \cite{FeAS,FeAs1}. Furthermore, the relationship between CDW order and superconductivity in the high $T_c$ cuprates has also received considerable interest \cite{competecompound,CuO}.  Therefore, it is desirable to characterize the pairing states of new CDW superconductors and to understand the interplay with CDW ordering.

The \emph{R}Pt$_2$\emph{X}$_2$ (\emph{R} = rare-earth/alkaline-earth element, \emph{X} = Si, As) compounds which crystallize in the CaBe$_2$Ge$_2$-type structure (space group \emph{P}4/\emph{nmm}) are a family of CDW superconductors with a variety of properties \cite{SrPt2As2,CDWseveralsamples,BaPt2AS22,pressure2}. SrPt$_2$As$_2$ undergoes a CDW transition at about 470~K and a superconducting transition at $T_c$~= 5.2~K \cite{SrPt2As2}, while BaPt$_2$As$_2$ exhibits CDW transition at 275~K and bulk superconductivity below 1.33~K \cite{BaPt2As2,BaPt2AS22}. Low-temperature measurements of the specific heat suggest that both SrPt$_2$As$_2$ and BaPt$_2$As$_2$ behave like \emph{s}-wave superconductors \cite{SrPtAsgap,BaPt2As2}. For BaPt$_2$As$_2$, a complex pressure-temperature phase diagram is obtained, showing multiple step like changes in $T_c$ concomitant with anomalies at elevated temperatures which may be associated with structural phase transitions \cite{BaPt2AS22}. Recently, both superconductivity and CDW order were found in the isostructural compound LaPt$_2$Si$_2$. The presence of CDW order was confirmed by selected area electron diffraction \cite{CDWseveralsamples} and x-ray diffraction results \cite{newr}. In contrast to BaPt$_2$As$_2$, LaPt$_2$Si$_2$ shows a superconducting dome near the critical pressure $P_c$ = 2.4 GPa where the CDW order suddenly vanishes \cite{pressure2}. However, this compound shows no evidence of the existence of a structural quantum critical point, and Fermi liquid behavior is preserved upon tuning with pressure.

In the superconducting state of LaPt$_2$Si$_2$, the specific heat shows an exponential temperature dependence, indicating fully gapped superconductivity \cite{singlecrystalandspecificheat}. Meanwhile, the superfluid density derived from muon-spin rotation ($\mu$SR) experiments could be fitted by either a two-gap \emph{s}-wave model or a \emph{d}-wave model \cite{mutigap-muSR}. Therefore, it is necessary to further study its gap structure using methods which are sensitive to the low-energy excitations. In this paper, we report measurements of the magnetic penetration depth using the tunnel-diode oscillator (TDO)-based method, as well as soft point-contact spectroscopy (soft-PCS) of single-crystalline LaPt$_2$Si$_2$. We find that the data from both  techniques are well described by a single-gap \emph{s}-wave model.

\section{EXPERIMENTAL DETAILS}

\begin{figure}
\includegraphics[angle=0,width=0.49\textwidth]{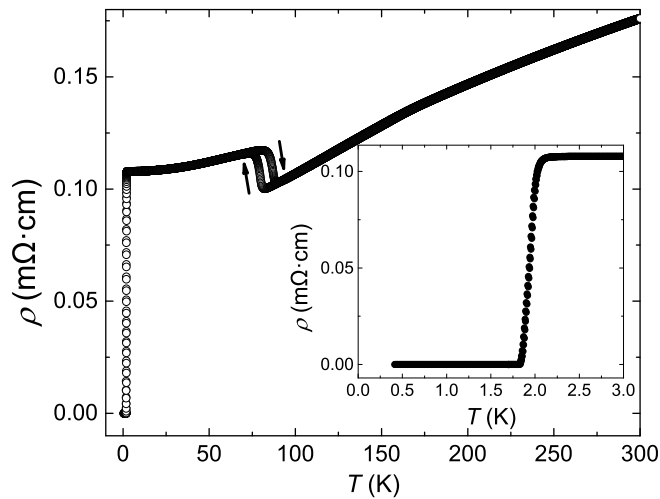}
\vspace{-12pt} \caption{\label{figure1}(Color online) Temperature dependence of the resistivity $\rho(T)$ of LaPt$_2$Si$_2$, measured upon both cooling and warming, as indicated by the arrows. The inset shows the low-temperature $\rho(T)$ near $\emph{T}_c$~=~1.8~K. }
\vspace{-12pt}
\end{figure}

Single crystals of LaPt$_2$Si$_2$ were synthesized using the Czochralski method, as described in Ref. \onlinecite{singlecrystalandspecificheat}. The electrical resistivity $\rho(T)$ was measured using a standard four-probe method, from 0.3~K to room temperature in a $^3$He cryostat. The London penetration depth shift $\Delta\lambda(T)$ was measured by using a TDO-based technique \cite{TDOdevice,TDO2}. The operating frequency of the TDO oscillator is about 7~MHz, with a noise level as low as 0.1~Hz. The samples used for the TDO measurements typically have dimensions of about 600~$\times$~300~$\times$~200~$\mu$m$^3$ and were mounted onto a sapphire rod. The small ac field generated by the coil, is about 20~mOe, which is much smaller than the lower critical field $H_{c1}$ of the sample, ensuring that the sample remains in the Meissner state. The change in the magnetic penetration depth is converted from the frequency shift by $\Delta\lambda(T)$~=$~G\Delta f(T)$, where \emph{G} is the calibration constant determined by the geometry of the sample and coil \cite{Gfactor}.

The soft-PCS technique was also applied to study its gap structure \cite{PCS,PCS2}, where a drop of silver conductive paint was placed between the surface of the sample and Au or Pt wires. The sample was polished to be mirror shining, and a 25-$\mu$m-diameter Pt wire was used to form point contacts. In the contact region, thousands of parallel conducting channels can be assumed to be on the sample surface. The bias-voltage dependance of the contact conductance \emph{G}(V) is measured in a quasi-four-probe configuration and recorded by the conventional lock-in technique, where the minus current and voltage probes bifurcate from the Au or Pt contact wire on the sample. A cryostat from Oxford Instruments with a $^3$He insert was used to cool the sample to a base temperature of 0.3~K.

\section{RESULTS}
\subsection{Penetration depth measurements}
Figure \ref{figure1} shows the electrical resistivity $\rho(T)$ of LaPt$_2$Si$_2$, which features a first order CDW transition, as indicated by the pronounced hysteresis loop. The transition temperature in our data is a little lower than that  reported for the polycrystalline samples \cite{CDWseveralsamples,thermaltransport}. The inset of Fig. \ref{figure1} shows the low-temperature $\rho(T)$, where a sharp superconducting transition onset at 2.1~K is observed, reaching zero resistivity at about $T_c$~=~1.8~K.

\begin{figure}
\includegraphics[angle=0,width=0.49\textwidth]{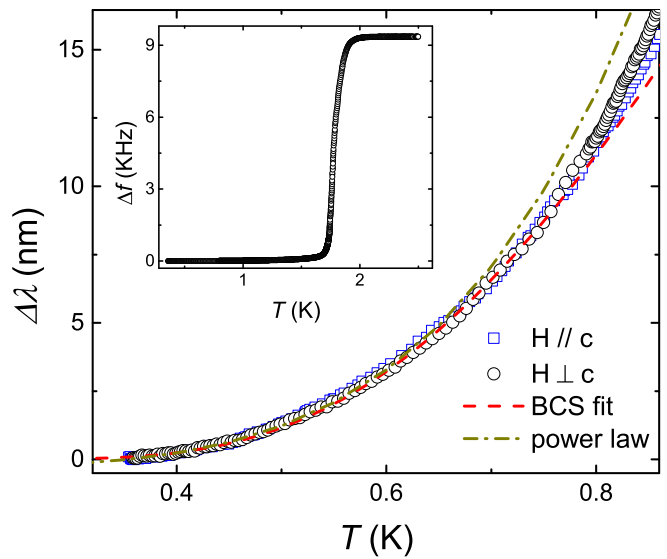}
\vspace{-12pt} \caption{\label{figure2}(Color online) $\Delta\lambda(T)$ of LaPt$_2$Si$_2$, where samples were measured with fields perpendicular and parallel to the \emph{c} axis. The dashed and dash-dotted lines represent fitting using an $s$-wave model and a power-law dependence, respectively. The inset shows $\Delta f(T)$ from 2.5 K down to 0.35~K, where there is a sharp superconducting transition around 1.8~K.}
\vspace{-12pt}
\end{figure}

Figure \ref{figure2} shows the temperature dependence of the London penetration depth shift $\Delta\lambda(T)$ of two LaPt$_2$Si$_2$ samples, which were measured with field perpendicular and parallel to the \emph{c} axis. Here the calibration factors are \emph{G}~=~22
~{\AA}/Hz and \emph{G}~=~18 {\AA}/Hz for $\emph{H} \parallel c$ and $\emph{H} \perp c$, respectively. The inset displays $\Delta f(T)$ with $\emph{H} \perp c$ from 2.5~K down to 0.35 K. It can be seen that there is a sharp superconducting transition in $\Delta f(T)$, with a midpoint of $T_c$ =~1.8 K. Upon lowering the temperature, $\Delta\lambda(T)$ flattens, indicating fully gapped superconductivity in LaPt$_2$Si$_2$. The similar behavior for the two field directions is consistent with an isotropic gap structure in LaPt$_2$Si$_2$. For a single-gap isotropic \emph{s}-wave superconductor, $\Delta\lambda(T)$ shows an exponentially activated temperature dependence for $T \ll T_c$,
 \begin{equation}
\Delta\lambda(T)=\lambda(0)\sqrt{\frac{\pi\Delta(0)}{2k_BT}}\textrm{exp}\left(-\frac{\Delta(0)}{k_BT}\right),
\label{equation1}
\end{equation}

\noindent where $\lambda(0)$ and $\Delta(0)$ are the magnetic penetration depth and superconducting gap magnitude at zero temperature, respectively. As displayed by the dashed line, our experimental data for two directions can be well described by an \emph{s}-wave model below $T_c$/3 with $\Delta(0)$~=~1.88$k_{B}T_c$, where $\lambda(0)$~=~279.3~nm is fixed from Ref. \onlinecite{mutigap-muSR}. If the data are fitted with power-law behavior $\Delta\lambda(T)$~$\propto$~$T^n$, from the base temperature up to 0.65~K, a large exponent of \emph{n}~=~4.7 is obtained, which indicates nodeless superconductivity. Therefore, our measurements clearly show that LaPt$_2$Si$_2$ is a fully gapped superconductor.

\begin{figure}
\includegraphics[angle=0,width=0.49\textwidth]{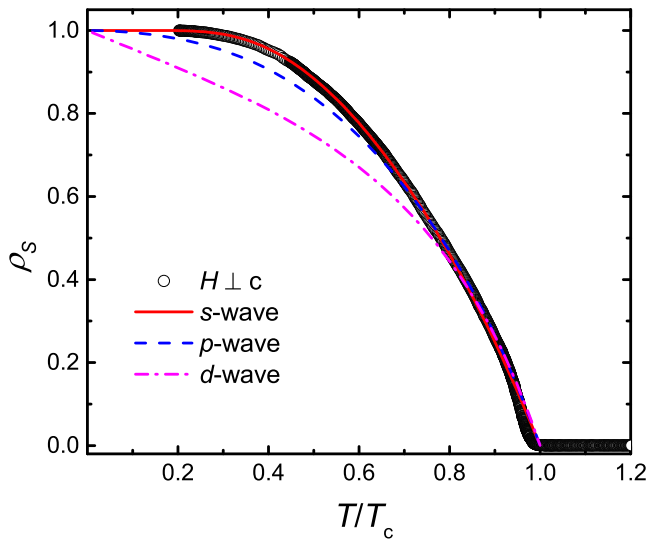}
\vspace{-12pt} \caption{\label{figure3}(Color online) Normalized superfluid density $\rho_s$ with $H \perp c$ as a function of the reduced temperature $T/T_c$. The solid, dashed, and dash-dotted lines here represent the fits for $s$-, $p$-, and $d$-wave models, respectively.}
\vspace{-12pt}
\end{figure}
To further investigate the superconducting gap structure of LaPt$_2$Si$_2$, we also analyzed the normalized superfluid density $\rho_s(T)$ derived by $\rho_s(T)~=~[\lambda(0)/\lambda(T)]^2$, using $\lambda(0)$~=~279.3~nm from Ref. \onlinecite{mutigap-muSR}. This quantity for $H \perp c$ is displayed as a function of the reduced temperature $T$/$T_c$ in Fig. \ref{figure3}. For a given gap function $\Delta_k$, the temperature dependence of $\rho_s(T)$ can be calculated by

\begin{equation}
\rho_{\rm s}(T) = 1 + 2 \left\langle\int_{\Delta_k}^{\infty}\frac{E{\rm d}E}{\sqrt{E^2-\Delta_k^2}}\frac{\partial f}{\partial E}\right\rangle_{\rm FS},
\label{equation2}
\end{equation}

\noindent where $f(E, T)=[1+{\rm exp}(E/k_BT)]^{-1}$ and $\Delta_k(T)=\Delta(T)$$g_k$ are the Fermi-Dirac distribution function and the superconducting gap function, respectively. The gap function is the product of an angle-dependent component $g_k$ and a temperature-dependent part $\Delta(T)$. $\Delta(T)$ was approximated using \cite{delta0}

\begin{equation}
\Delta(T)~=~\Delta(0){\rm tanh}\left\{1.82\left[1.018\left(T_c/T-1\right)\right]^{0.51}\right\}.
\label{equation3}
\end{equation}
\noindent

The results from fitting with different models are displayed in Fig. \ref{figure3}. In the case of a single-gap \emph{s}-wave model with $g_k$ = 1, this model can well describe the experimental data with a fitted value of $\Delta(0)$~=~0.31 meV, yielding $\Delta(0)$/$k_B$$\emph{T}_c$ = 2.0. The results from fitting with two nodal models are also displayed in Fig. \ref{figure3}. Here $g_k$ of sin~$\theta$ and cos~$2\phi$ were used, for a \emph{p}-wave model and \emph{d}-wave model respectively, where $\theta$ is the polar angle and $\phi$ is the azimuthal angle \cite{gk}. As shown in Fig. \ref{figure3}, both models display different behaviors at low temperatures in contrast to our data, which gives further evidence of nodeless superconductivity. These results therefore indicate that the superfluid density is well described by a single-gap \emph{s}-wave model. The deduced gap of 2.0$k_B$$T_c$, is close to the value 1.88$k_B$$T_c$ derived from fitting the low-temperature $\Delta\lambda(T)$. As these values are all larger than the value of 1.76$k_BT_c$ predicted from weakly coupled BCS theory, our results suggest a single-gap \emph{s}-wave superconductivity in LaPt$_2$Si$_2$ with moderately strong coupling.

\subsection{Point-contact spectroscopy measurements}

\begin{figure}
\includegraphics[angle=0,width=0.49\textwidth]{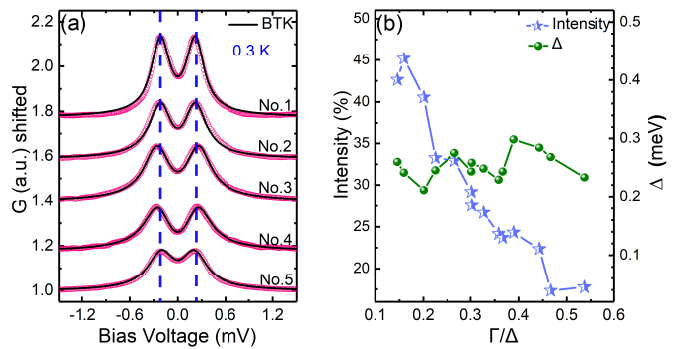}
\vspace{-12pt} \caption{\label{points}(Color online) (a) A set of representative point-contact conductance curves at 0.3 K in comparison with optimal BTK fitting curves (black solid lines). (b) A statistical plot of the conductance peak enhancement intensity (half-filled blue stars) and the extracted superconducting gap (olive green spheres) as a function of the smearing factor $\Gamma$ divided by the gap $\Delta$, $\Gamma/\Delta$, for different contacts. }
\vspace{-12pt}
\end{figure}

Dozens of soft point contacts on LaPt$_2$Si$_2$ single crystals were measured to study its gap structure, and several representative conductance curves from different contacts at a low temperature of~0.3 K are shown in Fig. \ref{points}(a). A common symmetric double-peak structure with the bias voltage around $\pm$~0.24~mV is observed for all curves, and the curves can be well fitted by the single-band \emph{s}-wave Blonder-Tinkham-Klapwijk (BTK) model \cite{BTK}. In particular, curves 3 and 5 in Fig. \ref{points}(a) show a very close match to the BTK fitting curves. All these conductance curves from different contacts consistently suggest a fully gapped behavior. A slight deviation at higher bias can be noticed for some curves due to the dip structure, which is generally believed to originate from the heating effect for the current across the contact. In such a case, the current can exceed the superconducting critical current, leading to a transition to the normal state \cite{CuBiSe,criticalcurrent}, and the obtained gap would be underestimated relative to the actual value. We note that our extracted gap values $\Delta$ are scattered only in a small range between 0.23 and 0.3~meV, yielding $2\Delta(0)/k_BT_c$ from 3.0 to 3.7 with $T_c$~=~1.9~K, close to the weak-coupling limit of 3.52. Meanwhile, the ratio of the quasiparticle smearing parameter $\Gamma$ to the gap value $\Delta$, $\Gamma/\Delta$, ranges from 0.15 to 0.55, indicating a wide range of the quasiparticle smearing with varying surface conditions of the contacts. If we summarize the obtained superconducting gap as a function of $\Gamma/\Delta$ for different contacts and plot them together with the relative Andreev reflection peak intensity as shown in Fig. \ref{points}(b), the gap value $\Delta$ does not change significantly with $\Gamma/\Delta$, while the peak intensity roughly shows a systematic decrease. It supports the idea that the obtained gap value is independent of the interface scattering with the spectroscopic nature of our soft point contacts.

\begin{figure}
\includegraphics[angle=0,width=0.49\textwidth]{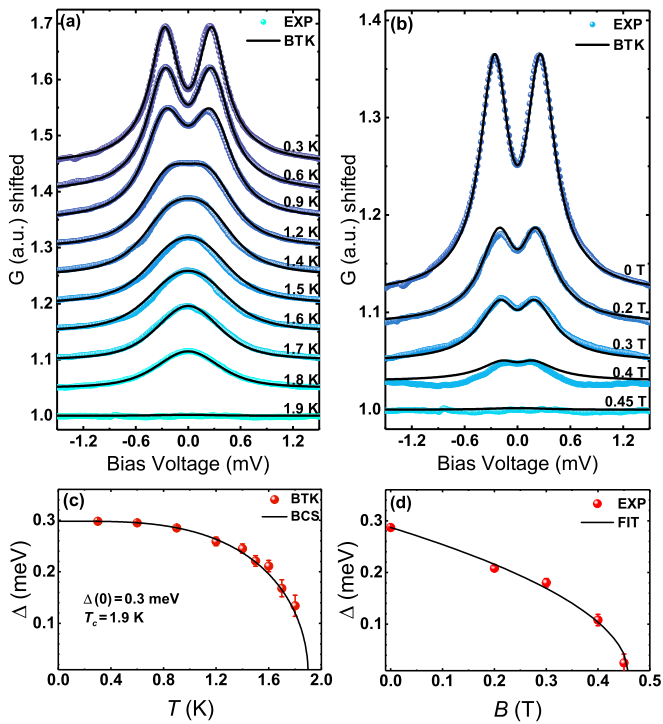}
\vspace{-12pt} \caption{\label{FigureGapevolution}(Color online) (a) and (b) The temperature and magnetic field evolution of the soft PCS on LaPt$_2$Si$_2$ in comparison with optimal BTK fitting curves. The curves are shifted for clarity, and the magnetic field is applied perpendicular to the sample surface. (c) and (d) The extracted superconducting gap value $\Delta$ from BTK fitting as a function of temperature and magnetic field, respectively. Solid lines in (c) and (d) show the BCS gap function and fitted curve. }
\vspace{-12pt}
\end{figure}

Figures \ref{FigureGapevolution}(a) and \ref{FigureGapevolution}(b) show the temperature and magnetic field dependence of the conductance curves \emph{G}(\emph{V}) in soft PCS, respectively, in comparison with the BTK fitting curves. With increasing temperature, the double peaks at 0.3 K get closer to each other and are smeared into a zero-bias peak at around 1.3 K and finally disappear at its superconducting transition temperature around 1.9 K. Figure \ref{FigureGapevolution}(c) shows the extracted gap values $\Delta$ from  the BTK fitting as a function of temperature, consistent with the BCS behavior. The extracted zero-temperature superconducting gap $\Delta(0)$ is determined to be 0.297(3) meV, yielding $2\Delta(0)/k_BT_c$~=~3.62(4), with $T_c$~=~1.9~K, slightly larger than the weak-coupling value of 3.52. On the other hand, the conductance curves as a function of magnetic field are shown in Fig. \ref{FigureGapevolution}(b) in comparison with the single-gap \emph{s}-wave fitting curves at \emph{T}~=~0.3 K. The Andreev reflection signal in the conductance curves is smeared with field up to around 0.45~T. The extracted gap values $\Delta(H)$ as a function of field are plotted in Fig. \ref{FigureGapevolution}(d) and follow the formula $\Delta(H)$ ~=~$\Delta(0)\sqrt{1-H/H_{c2}}$ \cite{MagneticFieldFormula}, consistent with a typical type-II superconductor in the field. The temperature and magnetic field evolution of soft PCS thus supports a single fully opened superconducting gap in LaPt$_2$Si$_2$ with $2\Delta_0/k_BT_c$~=~3.62(4).

\section{SUMMARY}

Our TDO and soft-PCS results for LaPt$_2$Si$_2$ both show that there is fully gapped \emph{s}-wave behavior, giving a consistent zero-temperature gap size of about 0.3 meV. For TDO measurements, an exponential behavior of $\Delta\lambda(T)$ was observed below $T_c/3$, and the superfluid density $\rho_s(T)$ could be well described by a single-gap \emph{s}-wave model. For soft-PCS measurements, the temperature and magnetic field evolution of the conductance curves can be well fitted by a single-gap \emph{s}-wave BTK model, and the extracted gap follows typical BCS behavior. However, we note that in Ref. \cite{mutigap-muSR}, it is reported that there is multi gap superconductivity in this material based on $\mu$SR measurements. One possible origin of this difference may be related to the difference between the single-crystal and polycrystalline samples.

\begin{acknowledgments}
This work was supported by the National Key R\&D Program of China (Grants No.~2017YFA0303100, No.~2016YFA0300202, No.~2016FYA0300402 and No.~2016YFA0401704), the National Natural Science Foundation of China (Grants No.~U1632275, No.~11974306, No.~11874320, No.~11674279, No.~12034017), and the Science Challenge Project of China (Grants No.~TZ2016004). X.L. would like to acknowledge support from the Zhejiang Provincial Natural Science Foundation of China (Grants No. LR18A04001). Z.H.  would like to acknowledge support from
SERB, India (Grant No.~CRG/2018/000220).

\end{acknowledgments}


\begin{thebibliography}{35}%
\makeatletter
\providecommand \@ifxundefined [1]{%
 \@ifx{#1\undefined}
}%
\providecommand \@ifnum [1]{%
 \ifnum #1\expandafter \@firstoftwo
 \else \expandafter \@secondoftwo
 \fi
}%
\providecommand \@ifx [1]{%
 \ifx #1\expandafter \@firstoftwo
 \else \expandafter \@secondoftwo
 \fi
}%
\providecommand \natexlab [1]{#1}%
\providecommand \enquote  [1]{``#1''}%
\providecommand \bibnamefont  [1]{#1}%
\providecommand \bibfnamefont [1]{#1}%
\providecommand \citenamefont [1]{#1}%
\providecommand \href@noop [0]{\@secondoftwo}%
\providecommand \href [0]{\begingroup \@sanitize@url \@href}%
\providecommand \@href[1]{\@@startlink{#1}\@@href}%
\providecommand \@@href[1]{\endgroup#1\@@endlink}%
\providecommand \@sanitize@url [0]{\catcode `\\12\catcode `\$12\catcode
  `\&12\catcode `\#12\catcode `\^12\catcode `\_12\catcode `\%12\relax}%
\providecommand \@@startlink[1]{}%
\providecommand \@@endlink[0]{}%
\providecommand \url  [0]{\begingroup\@sanitize@url \@url }%
\providecommand \@url [1]{\endgroup\@href {#1}{\urlprefix }}%
\providecommand \urlprefix  [0]{URL }%
\providecommand \Eprint [0]{\href }%
\providecommand \doibase [0]{https://doi.org/}%
\providecommand \selectlanguage [0]{\@gobble}%
\providecommand \bibinfo  [0]{\@secondoftwo}%
\providecommand \bibfield  [0]{\@secondoftwo}%
\providecommand \translation [1]{[#1]}%
\providecommand \BibitemOpen [0]{}%
\providecommand \bibitemStop [0]{}%
\providecommand \bibitemNoStop [0]{.\EOS\space}%
\providecommand \EOS [0]{\spacefactor3000\relax}%
\providecommand \BibitemShut  [1]{\csname bibitem#1\endcsname}%
\let\auto@bib@innerbib\@empty
\bibitem [{\citenamefont {Testardi}(1975)}]{A15}%
  \BibitemOpen
  \bibfield  {author} {\bibinfo {author} {\bibfnamefont {L.~R.}\ \bibnamefont
  {Testardi}},\ }\bibfield  {title} {\bibinfo {title} {Structural instability
  and superconductivity in $\mathrm{A}$-15 compounds},\ }\href
  {https://doi.org/10.1103/RevModPhys.47.637} {\bibfield  {journal} {\bibinfo
  {journal} {Rev. Mod. Phys.}\ }\textbf {\bibinfo {volume} {47}},\ \bibinfo
  {pages} {637} (\bibinfo {year} {1975})}\BibitemShut {NoStop}%
\bibitem [{\citenamefont {Wilson}\ \emph {et~al.}(1975)\citenamefont {Wilson},
  \citenamefont {Salvo},\ and\ \citenamefont {Mahajan}}]{CDW2}%
  \BibitemOpen
  \bibfield  {author} {\bibinfo {author} {\bibfnamefont {J.~A.}\ \bibnamefont
  {Wilson}}, \bibinfo {author} {\bibfnamefont {F.~J.~D.}\ \bibnamefont
  {Salvo}},\ and\ \bibinfo {author} {\bibfnamefont {S.}~\bibnamefont
  {Mahajan}},\ }\bibfield  {title} {\bibinfo {title} {Charge-density waves and
  superlattices in the metallic layered transition metal dichalcogenides},\
  }\href {https://doi.org/10.1080/00018737500101391} {\bibfield  {journal}
  {\bibinfo  {journal} {Adv. Phys.}\ }\textbf {\bibinfo {volume} {24}},\
  \bibinfo {pages} {117} (\bibinfo {year} {1975})}\BibitemShut {NoStop}%
\bibitem [{\citenamefont {Gabovich}\ \emph {et~al.}(2002)\citenamefont
  {Gabovich}, \citenamefont {Voitenko},\ and\ \citenamefont
  {Ausloos}}]{CDWandSC1}%
  \BibitemOpen
  \bibfield  {author} {\bibinfo {author} {\bibfnamefont {A.~M.}\ \bibnamefont
  {Gabovich}}, \bibinfo {author} {\bibfnamefont {A.~I.}\ \bibnamefont
  {Voitenko}},\ and\ \bibinfo {author} {\bibfnamefont {M.}~\bibnamefont
  {Ausloos}},\ }\bibfield  {title} {\bibinfo {title} {Charge- and spin-density
  waves in existing superconductors: Competition between cooper pairing and
  Peierls or excitonic instabilities},\ }\href
  {https://doi.org/https://doi.org/10.1016/S0370-1573(02)00029-7} {\bibfield
  {journal} {\bibinfo  {journal} {Phys. Rep.}\ }\textbf {\bibinfo {volume}
  {367}},\ \bibinfo {pages} {583 } (\bibinfo {year} {2002})}\BibitemShut
  {NoStop}%
\bibitem [{\citenamefont {Gabovich}\ \emph {et~al.}(2001)\citenamefont
  {Gabovich}, \citenamefont {Voitenko}, \citenamefont {Annett},\ and\
  \citenamefont {Ausloos}}]{CDWcompete}%
  \BibitemOpen
  \bibfield  {author} {\bibinfo {author} {\bibfnamefont {A.~M.}\ \bibnamefont
  {Gabovich}}, \bibinfo {author} {\bibfnamefont {A.~I.}\ \bibnamefont
  {Voitenko}}, \bibinfo {author} {\bibfnamefont {J.~F.}\ \bibnamefont
  {Annett}},\ and\ \bibinfo {author} {\bibfnamefont {M.}~\bibnamefont
  {Ausloos}},\ }\bibfield  {title} {\bibinfo {title} {Charge- and
  spin-density-wave superconductors},\ }\href
  {https://doi.org/10.1088/0953-2048/14/4/201} {\bibfield  {journal} {\bibinfo
  {journal} {Supercond. Sci. Technol.}\ }\textbf {\bibinfo {volume} {14}},\
  \bibinfo {pages} {R1} (\bibinfo {year} {2001})}\BibitemShut {NoStop}%
\bibitem [{\citenamefont {Monteverde}\ \emph {et~al.}(2013)\citenamefont
  {Monteverde}, \citenamefont {Lorenzana}, \citenamefont {Monceau},\ and\
  \citenamefont {N\'u\~nez Regueiro}}]{TaS}%
  \BibitemOpen
  \bibfield  {author} {\bibinfo {author} {\bibfnamefont {M.}~\bibnamefont
  {Monteverde}}, \bibinfo {author} {\bibfnamefont {J.}~\bibnamefont
  {Lorenzana}}, \bibinfo {author} {\bibfnamefont {P.}~\bibnamefont {Monceau}},\
  and\ \bibinfo {author} {\bibfnamefont {M.}~\bibnamefont {N\'u\~nez--Regueiro}},\ }\bibfield  {title} {\bibinfo {title} {Quantum critical point
  and superconducting dome in the pressure phase diagram of
  $o$-$\mathrm{TaS}_{3}$},\ }\href {https://doi.org/10.1103/PhysRevB.88.180504}
  {\bibfield  {journal} {\bibinfo  {journal} {Phys. Rev. B}\ }\textbf {\bibinfo
  {volume} {88}},\ \bibinfo {pages} {180504(R)} (\bibinfo {year}
  {2013})}\BibitemShut {NoStop}%
\bibitem [{\citenamefont {Morosan}\ \emph {et~al.}(2006)\citenamefont
  {Morosan}, \citenamefont {Zandbergen}, \citenamefont {Dennis}, \citenamefont
  {Bos}, \citenamefont {Onose}, \citenamefont {Klimczuk}, \citenamefont
  {Ramirez}, \citenamefont {Ong},\ and\ \citenamefont {Cava}}]{CuTiSe2}%
  \BibitemOpen
  \bibfield  {author} {\bibinfo {author} {\bibfnamefont {E.}~\bibnamefont
  {Morosan}}, \bibinfo {author} {\bibfnamefont {H.~W.}\ \bibnamefont
  {Zandbergen}}, \bibinfo {author} {\bibfnamefont {B.~S.}\ \bibnamefont
  {Dennis}}, \bibinfo {author} {\bibfnamefont {J.~W.~G.}\ \bibnamefont {Bos}},
  \bibinfo {author} {\bibfnamefont {Y.}~\bibnamefont {Onose}}, \bibinfo
  {author} {\bibfnamefont {T.}~\bibnamefont {Klimczuk}}, \bibinfo {author}
  {\bibfnamefont {A.~P.}\ \bibnamefont {Ramirez}}, \bibinfo {author}
  {\bibfnamefont {N.~P.}\ \bibnamefont {Ong}},\ and\ \bibinfo {author}
  {\bibfnamefont {R.~J.}\ \bibnamefont {Cava}},\ }\bibfield  {title} {\bibinfo
  {title} {Superconductivity in $\mathrm{Cu}_{x}\mathrm{TiSe}_{2}$},\ }\href
  {https://doi.org/10.1038/nphys360} {\bibfield  {journal} {\bibinfo  {journal}
  {Nat. Phys.}\ }\textbf {\bibinfo {volume} {2}},\ \bibinfo {pages} {544}
  (\bibinfo {year} {2006})}\BibitemShut {NoStop}%
\bibitem [{\citenamefont {Gruner}\ \emph {et~al.}(2017)\citenamefont {Gruner},
  \citenamefont {Jang}, \citenamefont {Huesges}, \citenamefont {Cardoso-Gil},
  \citenamefont {Fecher}, \citenamefont {Koza}, \citenamefont {Stockert},
  \citenamefont {Mackenzie}, \citenamefont {Brando},\ and\ \citenamefont
  {Geibel}}]{lpi1}%
  \BibitemOpen
  \bibfield  {author} {\bibinfo {author} {\bibfnamefont {T.}~\bibnamefont
  {Gruner}}, \bibinfo {author} {\bibfnamefont {D.}~\bibnamefont {Jang}},
  \bibinfo {author} {\bibfnamefont {Z.}~\bibnamefont {Huesges}}, \bibinfo
  {author} {\bibfnamefont {R.}~\bibnamefont {Cardoso-Gil}}, \bibinfo {author}
  {\bibfnamefont {G.~H.}\ \bibnamefont {Fecher}}, \bibinfo {author}
  {\bibfnamefont {M.~M.}\ \bibnamefont {Koza}}, \bibinfo {author}
  {\bibfnamefont {O.}~\bibnamefont {Stockert}}, \bibinfo {author}
  {\bibfnamefont {A.~P.}\ \bibnamefont {Mackenzie}}, \bibinfo {author}
  {\bibfnamefont {M.}~\bibnamefont {Brando}},\ and\ \bibinfo {author}
  {\bibfnamefont {C.}~\bibnamefont {Geibel}},\ }\bibfield  {title} {\bibinfo
  {title} {Charge density wave quantum critical point with strong enhancement
  of superconductivity},\ }\href {https://doi.org/10.1038/nphys4191} {\bibfield
   {journal} {\bibinfo  {journal} {Nat. Phys.}\ }\textbf {\bibinfo {volume}
  {13}},\ \bibinfo {pages} {967} (\bibinfo {year} {2017})}\BibitemShut
  {NoStop}%
\bibitem [{\citenamefont {Goh}\ \emph {et~al.}(2015)\citenamefont {Goh},
  \citenamefont {Tompsett}, \citenamefont {Saines}, \citenamefont {Chang},
  \citenamefont {Matsumoto}, \citenamefont {Imai}, \citenamefont {Yoshimura},\
  and\ \citenamefont {Grosche}}]{CaRhSn}%
  \BibitemOpen
  \bibfield  {author} {\bibinfo {author} {\bibfnamefont {S.~K.}\ \bibnamefont
  {Goh}}, \bibinfo {author} {\bibfnamefont {D.~A.}\ \bibnamefont {Tompsett}},
  \bibinfo {author} {\bibfnamefont {P.~J.}\ \bibnamefont {Saines}}, \bibinfo
  {author} {\bibfnamefont {H.~C.}\ \bibnamefont {Chang}}, \bibinfo {author}
  {\bibfnamefont {T.}~\bibnamefont {Matsumoto}}, \bibinfo {author}
  {\bibfnamefont {M.}~\bibnamefont {Imai}}, \bibinfo {author} {\bibfnamefont
  {K.}~\bibnamefont {Yoshimura}},\ and\ \bibinfo {author} {\bibfnamefont
  {F.~M.}\ \bibnamefont {Grosche}},\ }\bibfield  {title} {\bibinfo {title}
  {Ambient pressure structural quantum critical point in the phase diagram of
  ${(\mathrm{Ca}_{x}\mathrm{Sr}_{1\ensuremath{-}x})}_{3}\mathrm{Rh}_{4}\mathrm{Sn}_{13}$},\
  }\href {https://doi.org/10.1103/PhysRevLett.114.097002} {\bibfield  {journal}
  {\bibinfo  {journal} {Phys. Rev. Lett.}\ }\textbf {\bibinfo {volume} {114}},\
  \bibinfo {pages} {097002} (\bibinfo {year} {2015})}\BibitemShut {NoStop}%
\bibitem [{\citenamefont {Mathur}\ \emph {et~al.}(1998)\citenamefont {Mathur},
  \citenamefont {Grosche}, \citenamefont {Julian}, \citenamefont {Walker},
  \citenamefont {Freye}, \citenamefont {Haselwimmer},\ and\ \citenamefont
  {Lonzarich}}]{CeIn3}%
  \BibitemOpen
  \bibfield  {author} {\bibinfo {author} {\bibfnamefont {N.~D.}\ \bibnamefont
  {Mathur}}, \bibinfo {author} {\bibfnamefont {F.~M.}\ \bibnamefont {Grosche}},
  \bibinfo {author} {\bibfnamefont {S.~R.}\ \bibnamefont {Julian}}, \bibinfo
  {author} {\bibfnamefont {I.~R.}\ \bibnamefont {Walker}}, \bibinfo {author}
  {\bibfnamefont {D.~M.}\ \bibnamefont {Freye}}, \bibinfo {author}
  {\bibfnamefont {R.~K.~W.}\ \bibnamefont {Haselwimmer}},\ and\ \bibinfo
  {author} {\bibfnamefont {G.~G.}\ \bibnamefont {Lonzarich}},\ }\bibfield
  {title} {\bibinfo {title} {Magnetically mediated superconductivity in heavy
  fermion compounds},\ }\href {https://doi.org/10.1038/27838} {\bibfield
  {journal} {\bibinfo  {journal} {Nature (London)}\ }\textbf {\bibinfo {volume} {394}},\
  \bibinfo {pages} {39} (\bibinfo {year} {1998})}\BibitemShut {NoStop}%
\bibitem [{\citenamefont {Yuan}\ \emph {et~al.}(2003)\citenamefont {Yuan},
  \citenamefont {Grosche}, \citenamefont {Deppe}, \citenamefont {Geibel},
  \citenamefont {Sparn},\ and\ \citenamefont {Steglich}}]{CCS}%
  \BibitemOpen
  \bibfield  {author} {\bibinfo {author} {\bibfnamefont {H.~Q.}\ \bibnamefont
  {Yuan}}, \bibinfo {author} {\bibfnamefont {F.~M.}\ \bibnamefont {Grosche}},
  \bibinfo {author} {\bibfnamefont {M.}~\bibnamefont {Deppe}}, \bibinfo
  {author} {\bibfnamefont {C.}~\bibnamefont {Geibel}}, \bibinfo {author}
  {\bibfnamefont {G.}~\bibnamefont {Sparn}},\ and\ \bibinfo {author}
  {\bibfnamefont {F.}~\bibnamefont {Steglich}},\ }\bibfield  {title} {\bibinfo
  {title} {Observation of two distinct superconducting phases in
  $\mathrm{CeCu}_2\mathrm{Si}_2$},\ }\href
  {https://doi.org/10.1126/science.1091648} {\bibfield  {journal} {\bibinfo
  {journal} {Science}\ }\textbf {\bibinfo {volume} {302}},\ \bibinfo {pages}
  {2104} (\bibinfo {year} {2003})}\BibitemShut {NoStop}%
\bibitem [{\citenamefont {de~la Cruz}\ \emph {et~al.}(2008)\citenamefont {de~la
  Cruz}, \citenamefont {Huang}, \citenamefont {lynn}, \citenamefont {Li},
  \citenamefont {Ratcliff~II}, \citenamefont {Zarestky}, \citenamefont {Mook},
  \citenamefont {Chen}, \citenamefont {Luo}, \citenamefont {Wang},\ and\
  \citenamefont {Dai}}]{FeAS}%
  \BibitemOpen
  \bibfield  {author} {\bibinfo {author} {\bibfnamefont {C.}~\bibnamefont
  {de~la Cruz}}, \bibinfo {author} {\bibfnamefont {Q.}~\bibnamefont {Huang}},
  \bibinfo {author} {\bibfnamefont {J.~W.}\ \bibnamefont {Lynn}}, \bibinfo
  {author} {\bibfnamefont {J.}~\bibnamefont {Li}}, \bibinfo {author}
  {\bibfnamefont {W.}~\bibnamefont {Ratcliff~II}}, \bibinfo {author}
  {\bibfnamefont {J.~L.}\ \bibnamefont {Zarestky}}, \bibinfo {author}
  {\bibfnamefont {H.~A.}\ \bibnamefont {Mook}}, \bibinfo {author}
  {\bibfnamefont {G.~F.}\ \bibnamefont {Chen}}, \bibinfo {author}
  {\bibfnamefont {J.~L.}\ \bibnamefont {Luo}}, \bibinfo {author} {\bibfnamefont
  {N.~L.}\ \bibnamefont {Wang}},\ and\ \bibinfo {author} {\bibfnamefont
  {P.}~\bibnamefont {Dai}},\ }\bibfield  {title} {\bibinfo {title} {Magnetic
  order close to superconductivity in the iron-based layered
  $\mathrm{LaO}_{1-x}\mathrm{F}_{x}\mathrm{FeAs}$ systems},\ }\href
  {https://doi.org/10.1038/nature07057} {\bibfield  {journal} {\bibinfo
  {journal} {Nature (London)}\ }\textbf {\bibinfo {volume} {453}},\ \bibinfo {pages}
  {899} (\bibinfo {year} {2008})}\BibitemShut {NoStop}%
\bibitem [{\citenamefont {Ni}\ \emph {et~al.}(2008)\citenamefont {Ni},
  \citenamefont {Tillman}, \citenamefont {Yan}, \citenamefont {Kracher},
  \citenamefont {Hannahs}, \citenamefont {Bud'ko},\ and\ \citenamefont
  {Canfield}}]{FeAs1}%
  \BibitemOpen
  \bibfield  {author} {\bibinfo {author} {\bibfnamefont {N.}~\bibnamefont
  {Ni}}, \bibinfo {author} {\bibfnamefont {M.~E.}\ \bibnamefont {Tillman}},
  \bibinfo {author} {\bibfnamefont {J.~Q.}\ \bibnamefont {Yan}}, \bibinfo
  {author} {\bibfnamefont {A.}~\bibnamefont {Kracher}}, \bibinfo {author}
  {\bibfnamefont {S.~T.}\ \bibnamefont {Hannahs}}, \bibinfo {author}
  {\bibfnamefont {S.~L.}\ \bibnamefont {Bud'ko}},\ and\ \bibinfo {author}
  {\bibfnamefont {P.~C.}\ \bibnamefont {Canfield}},\ }\bibfield  {title}
  {\bibinfo {title} {Effects of $\mathrm{C}$o substitution on thermodynamic and transport
  properties and anisotropic ${H}_{c2}$ in
  $\mathrm{Ba}{({\mathrm{Fe}}_{1\ensuremath{-}x}{\mathrm{Co}}_{x})}_{2}\mathrm{As}_{2}$
  single crystals},\ }\href {https://doi.org/10.1103/PhysRevB.78.214515}
  {\bibfield  {journal} {\bibinfo  {journal} {Phys. Rev. B}\ }\textbf {\bibinfo
  {volume} {78}},\ \bibinfo {pages} {214515} (\bibinfo {year}
  {2008})}\BibitemShut {NoStop}%
\bibitem [{\citenamefont {Chang}\ \emph {et~al.}(2012)\citenamefont {Chang},
  \citenamefont {Blackburn}, \citenamefont {Holmes}, \citenamefont
  {Christensen}, \citenamefont {Larsen}, \citenamefont {Mesot}, \citenamefont
  {Liang}, \citenamefont {Bonn}, \citenamefont {Hardy}, \citenamefont
  {Watenphul}, \citenamefont {Zimmermann}, \citenamefont {Forgan},\ and\
  \citenamefont {Hayden}}]{competecompound}%
  \BibitemOpen
  \bibfield  {author} {\bibinfo {author} {\bibfnamefont {J.}~\bibnamefont
  {Chang}}, \bibinfo {author} {\bibfnamefont {E.}~\bibnamefont {Blackburn}},
  \bibinfo {author} {\bibfnamefont {A.~T.}\ \bibnamefont {Holmes}}, \bibinfo
  {author} {\bibfnamefont {N.~B.}\ \bibnamefont {Christensen}}, \bibinfo
  {author} {\bibfnamefont {J.}~\bibnamefont {Larsen}}, \bibinfo {author}
  {\bibfnamefont {J.}~\bibnamefont {Mesot}}, \bibinfo {author} {\bibfnamefont
  {R.}~\bibnamefont {Liang}}, \bibinfo {author} {\bibfnamefont {D.~A.}\
  \bibnamefont {Bonn}}, \bibinfo {author} {\bibfnamefont {W.~N.}\ \bibnamefont
  {Hardy}}, \bibinfo {author} {\bibfnamefont {A.}~\bibnamefont {Watenphul}},
  \bibinfo {author} {\bibfnamefont {M.~v.}\ \bibnamefont {Zimmermann}},
  \bibinfo {author} {\bibfnamefont {E.~M.}\ \bibnamefont {Forgan}},\ and\
  \bibinfo {author} {\bibfnamefont {S.~M.}\ \bibnamefont {Hayden}},\ }\bibfield
   {title} {\bibinfo {title} {Direct observation of competition between
  superconductivity and charge density wave order in
  $\mathrm{YBa}_2\mathrm{Cu}_3\mathrm{O}_{6.67}$},\ }\href
  {https://doi.org/10.1038/nphys2456} {\bibfield  {journal} {\bibinfo
  {journal} {Nat. Phys.}\ }\textbf {\bibinfo {volume} {8}},\ \bibinfo {pages}
  {871} (\bibinfo {year} {2012})}\BibitemShut {NoStop}%
\bibitem [{\citenamefont {da~Silva~Neto}\ \emph {et~al.}(2014)\citenamefont
  {da~Silva~Neto}, \citenamefont {Aynajian}, \citenamefont {Frano},
  \citenamefont {Comin}, \citenamefont {Schierle}, \citenamefont {Weschke},
  \citenamefont {Gyenis}, \citenamefont {Wen}, \citenamefont {Schneeloch},
  \citenamefont {Xu}, \citenamefont {Ono}, \citenamefont {Gu}, \citenamefont
  {Le~Tacon},\ and\ \citenamefont {Yazdani}}]{CuO}%
  \BibitemOpen
  \bibfield  {author} {\bibinfo {author} {\bibfnamefont {E.~H.}\ \bibnamefont
  {da~Silva~Neto}}, \bibinfo {author} {\bibfnamefont {P.}~\bibnamefont
  {Aynajian}}, \bibinfo {author} {\bibfnamefont {A.}~\bibnamefont {Frano}},
  \bibinfo {author} {\bibfnamefont {R.}~\bibnamefont {Comin}}, \bibinfo
  {author} {\bibfnamefont {E.}~\bibnamefont {Schierle}}, \bibinfo {author}
  {\bibfnamefont {E.}~\bibnamefont {Weschke}}, \bibinfo {author} {\bibfnamefont
  {A.}~\bibnamefont {Gyenis}}, \bibinfo {author} {\bibfnamefont
  {J.}~\bibnamefont {Wen}}, \bibinfo {author} {\bibfnamefont {J.}~\bibnamefont
  {Schneeloch}}, \bibinfo {author} {\bibfnamefont {Z.}~\bibnamefont {Xu}},
  \bibinfo {author} {\bibfnamefont {S.}~\bibnamefont {Ono}}, \bibinfo {author}
  {\bibfnamefont {G.}~\bibnamefont {Gu}}, \bibinfo {author} {\bibfnamefont
  {M.}~\bibnamefont {Le~Tacon}},\ and\ \bibinfo {author} {\bibfnamefont
  {A.}~\bibnamefont {Yazdani}},\ }\bibfield  {title} {\bibinfo {title}
  {Ubiquitous interplay between charge ordering and high-temperature
  superconductivity in cuprates},\ }\href
  {https://doi.org/10.1126/science.1243479} {\bibfield  {journal} {\bibinfo
  {journal} {Science}\ }\textbf {\bibinfo {volume} {343}},\ \bibinfo {pages}
  {393} (\bibinfo {year} {2014})}\BibitemShut {NoStop}%
\bibitem [{\citenamefont {Kudo}\ \emph {et~al.}(2010)\citenamefont {Kudo},
  \citenamefont {Nishikubo},\ and\ \citenamefont {Nohara}}]{SrPt2As2}%
  \BibitemOpen
  \bibfield  {author} {\bibinfo {author} {\bibfnamefont {K.}~\bibnamefont
  {Kudo}}, \bibinfo {author} {\bibfnamefont {Y.}~\bibnamefont {Nishikubo}},\
  and\ \bibinfo {author} {\bibfnamefont {M.}~\bibnamefont {Nohara}},\
  }\bibfield  {title} {\bibinfo {title} {Coexistence of superconductivity and
  charge density wave in $\mathrm{SrPt}_2\mathrm{As}_2$},\ }\href
  {https://doi.org/10.1143/JPSJ.79.123710} {\bibfield  {journal} {\bibinfo
  {journal} {J. Phys. Soc. Jpn.}\ }\textbf {\bibinfo {volume} {79}},\ \bibinfo
  {pages} {123710} (\bibinfo {year} {2010})}\BibitemShut {NoStop}%
\bibitem [{\citenamefont {Nagano}\ \emph {et~al.}(2013)\citenamefont {Nagano},
  \citenamefont {Araoka}, \citenamefont {Mitsuda}, \citenamefont {Yayama},
  \citenamefont {Wada}, \citenamefont {Ichihara}, \citenamefont {Isobe},\ and\
  \citenamefont {Ueda}}]{CDWseveralsamples}%
  \BibitemOpen
  \bibfield  {author} {\bibinfo {author} {\bibfnamefont {Y.}~\bibnamefont
  {Nagano}}, \bibinfo {author} {\bibfnamefont {N.}~\bibnamefont {Araoka}},
  \bibinfo {author} {\bibfnamefont {A.}~\bibnamefont {Mitsuda}}, \bibinfo
  {author} {\bibfnamefont {H.}~\bibnamefont {Yayama}}, \bibinfo {author}
  {\bibfnamefont {H.}~\bibnamefont {Wada}}, \bibinfo {author} {\bibfnamefont
  {M.}~\bibnamefont {Ichihara}}, \bibinfo {author} {\bibfnamefont
  {M.}~\bibnamefont {Isobe}},\ and\ \bibinfo {author} {\bibfnamefont
  {Y.}~\bibnamefont {Ueda}},\ }\bibfield  {title} {\bibinfo {title} {Charge
  density wave and superconductivity of $\mathrm{RPt_{2}Si_{2}}$ ({R = Y, {L}a,
  Nd, and Lu})},\ }\href {https://doi.org/10.7566/JPSJ.82.064715} {\bibfield
  {journal} {\bibinfo  {journal} {J. Phys. Soc. Jpn.}\ }\textbf {\bibinfo
  {volume} {82}},\ \bibinfo {pages} {064715} (\bibinfo {year}
  {2013})}\BibitemShut {NoStop}%
\bibitem [{\citenamefont {Guo}\ \emph {et~al.}(2016)\citenamefont {Guo},
  \citenamefont {Jiang}, \citenamefont {Smidman}, \citenamefont {Han},
  \citenamefont {Malliakas}, \citenamefont {Shen}, \citenamefont {Wang},
  \citenamefont {Chen}, \citenamefont {Lu}, \citenamefont {Kanatzidis},\ and\
  \citenamefont {Yuan}}]{BaPt2AS22}%
  \BibitemOpen
  \bibfield  {author} {\bibinfo {author} {\bibfnamefont {C.~Y.}\ \bibnamefont
  {Guo}}, \bibinfo {author} {\bibfnamefont {W.~B.}\ \bibnamefont {Jiang}},
  \bibinfo {author} {\bibfnamefont {M.}~\bibnamefont {Smidman}}, \bibinfo
  {author} {\bibfnamefont {F.}~\bibnamefont {Han}}, \bibinfo {author}
  {\bibfnamefont {C.~D.}\ \bibnamefont {Malliakas}}, \bibinfo {author}
  {\bibfnamefont {B.}~\bibnamefont {Shen}}, \bibinfo {author} {\bibfnamefont
  {Y.~F.}\ \bibnamefont {Wang}}, \bibinfo {author} {\bibfnamefont
  {Y.}~\bibnamefont {Chen}}, \bibinfo {author} {\bibfnamefont {X.}~\bibnamefont
  {Lu}}, \bibinfo {author} {\bibfnamefont {M.~G.}\ \bibnamefont {Kanatzidis}},\
  and\ \bibinfo {author} {\bibfnamefont {H.~Q.}\ \bibnamefont {Yuan}},\
  }\bibfield  {title} {\bibinfo {title} {Superconductivity and multiple
  pressure-induced phases in $\mathrm{BaPt}_{2}\mathrm{As}_{2}$},\ }\href
  {https://doi.org/10.1103/PhysRevB.94.184506} {\bibfield  {journal} {\bibinfo
  {journal} {Phys. Rev. B}\ }\textbf {\bibinfo {volume} {94}},\ \bibinfo
  {pages} {184506} (\bibinfo {year} {2016})}\BibitemShut {NoStop}%
\bibitem [{\citenamefont {Shen}\ \emph {et~al.}(2020)\citenamefont {Shen},
  \citenamefont {Du}, \citenamefont {Li}, \citenamefont {Thamizhavel},
  \citenamefont {Smidman}, \citenamefont {Nie}, \citenamefont {Luo},
  \citenamefont {Le}, \citenamefont {Hossain},\ and\ \citenamefont
  {Yuan}}]{pressure2}%
  \BibitemOpen
  \bibfield  {author} {\bibinfo {author} {\bibfnamefont {B.}~\bibnamefont
  {Shen}}, \bibinfo {author} {\bibfnamefont {F.}~\bibnamefont {Du}}, \bibinfo
  {author} {\bibfnamefont {R.}~\bibnamefont {Li}}, \bibinfo {author}
  {\bibfnamefont {A.}~\bibnamefont {Thamizhavel}}, \bibinfo {author}
  {\bibfnamefont {M.}~\bibnamefont {Smidman}}, \bibinfo {author} {\bibfnamefont
  {Z.~Y.}\ \bibnamefont {Nie}}, \bibinfo {author} {\bibfnamefont {S.~S.}\
  \bibnamefont {Luo}}, \bibinfo {author} {\bibfnamefont {T.}~\bibnamefont
  {Le}}, \bibinfo {author} {\bibfnamefont {Z.}~\bibnamefont {Hossain}},\ and\
  \bibinfo {author} {\bibfnamefont {H.~Q.}\ \bibnamefont {Yuan}},\ }\bibfield
  {title} {\bibinfo {title} {Evolution of charge density wave order and
  superconductivity under pressure in $\mathrm{LaPt}_2\mathrm{Si}_2$},\ }\href
  {https://doi.org/10.1103/PhysRevB.101.144501} {\bibfield  {journal} {\bibinfo
   {journal} {Phys. Rev. B}\ }\textbf {\bibinfo {volume} {101}},\ \bibinfo
  {pages} {144501} (\bibinfo {year} {2020})}\BibitemShut {NoStop}%
\bibitem [{\citenamefont {Jiang}\ \emph {et~al.}(2014)\citenamefont {Jiang},
  \citenamefont {Guo}, \citenamefont {Weng}, \citenamefont {Wang},
  \citenamefont {Chen}, \citenamefont {Chen}, \citenamefont {Pang},
  \citenamefont {Shang}, \citenamefont {Lu},\ and\ \citenamefont
  {Yuan}}]{BaPt2As2}%
  \BibitemOpen
  \bibfield  {author} {\bibinfo {author} {\bibfnamefont {W.~B.}\ \bibnamefont
  {Jiang}}, \bibinfo {author} {\bibfnamefont {C.~Y.}\ \bibnamefont {Guo}},
  \bibinfo {author} {\bibfnamefont {Z.~F.}\ \bibnamefont {Weng}}, \bibinfo
  {author} {\bibfnamefont {Y.~F.}\ \bibnamefont {Wang}}, \bibinfo {author}
  {\bibfnamefont {Y.~H.}\ \bibnamefont {Chen}}, \bibinfo {author}
  {\bibfnamefont {Y.}~\bibnamefont {Chen}}, \bibinfo {author} {\bibfnamefont
  {G.~M.}\ \bibnamefont {Pang}}, \bibinfo {author} {\bibfnamefont
  {T.}~\bibnamefont {Shang}}, \bibinfo {author} {\bibfnamefont
  {X.}~\bibnamefont {Lu}},\ and\ \bibinfo {author} {\bibfnamefont {H.~Q.}\
  \bibnamefont {Yuan}},\ }\bibfield  {title} {\bibinfo {title}
  {Superconductivity and structural distortion in
  $\mathrm{BaPt}_2\mathrm{As}_2$},\ }\href
  {https://doi.org/10.1088/0953-8984/27/2/022202} {\bibfield  {journal}
  {\bibinfo  {journal} {J. Phys.: Condens. Matter}\ }\textbf {\bibinfo {volume}
  {27}},\ \bibinfo {pages} {022202} (\bibinfo {year} {2014})}\BibitemShut
  {NoStop}%
\bibitem [{\citenamefont {Xu}\ \emph {et~al.}(2013)\citenamefont {Xu},
  \citenamefont {Chen}, \citenamefont {Jiao}, \citenamefont {Chen},
  \citenamefont {Niu}, \citenamefont {Li}, \citenamefont {Yang}, \citenamefont
  {Bangura}, \citenamefont {Ye}, \citenamefont {Cao}, \citenamefont {Dai},
  \citenamefont {Cao},\ and\ \citenamefont {Hussey}}]{SrPtAsgap}%
  \BibitemOpen
  \bibfield  {author} {\bibinfo {author} {\bibfnamefont {X.}~\bibnamefont
  {Xu}}, \bibinfo {author} {\bibfnamefont {B.}~\bibnamefont {Chen}}, \bibinfo
  {author} {\bibfnamefont {W.~H.}\ \bibnamefont {Jiao}}, \bibinfo {author}
  {\bibfnamefont {B.}~\bibnamefont {Chen}}, \bibinfo {author} {\bibfnamefont
  {C.~Q.}\ \bibnamefont {Niu}}, \bibinfo {author} {\bibfnamefont {Y.~K.}\
  \bibnamefont {Li}}, \bibinfo {author} {\bibfnamefont {J.~H.}\ \bibnamefont
  {Yang}}, \bibinfo {author} {\bibfnamefont {A.~F.}\ \bibnamefont {Bangura}},
  \bibinfo {author} {\bibfnamefont {Q.~L.}\ \bibnamefont {Ye}}, \bibinfo
  {author} {\bibfnamefont {C.}~\bibnamefont {Cao}}, \bibinfo {author}
  {\bibfnamefont {J.~H.}\ \bibnamefont {Dai}}, \bibinfo {author} {\bibfnamefont
  {G.}~\bibnamefont {Cao}},\ and\ \bibinfo {author} {\bibfnamefont {N.~E.}\
  \bibnamefont {Hussey}},\ }\bibfield  {title} {\bibinfo {title} {Evidence for
  two energy gaps and Fermi liquid behavior in the
  $\mathrm{SrPt}_{2}\mathrm{As}_{2}$ superconductor},\ }\href
  {https://doi.org/10.1103/PhysRevB.87.224507} {\bibfield  {journal} {\bibinfo
  {journal} {Phys. Rev. B}\ }\textbf {\bibinfo {volume} {87}},\ \bibinfo
  {pages} {224507} (\bibinfo {year} {2013})}\BibitemShut {NoStop}%
\bibitem [{\citenamefont {Falkowski}\ \emph {et~al.}(2019)\citenamefont
  {Falkowski}, \citenamefont {Dole\ifmmode~\check{z}\else \v{z}\fi{}al},
  \citenamefont {Andreev}, \citenamefont {Duverger-N\'edellec},\ and\
  \citenamefont {Havela}}]{newr}%
  \BibitemOpen
  \bibfield  {author} {\bibinfo {author} {\bibfnamefont {M.}~\bibnamefont
  {Falkowski}}, \bibinfo {author} {\bibfnamefont {P.}~\bibnamefont
  {Dole\ifmmode~\check{z}\else \v{z}\fi{}al}}, \bibinfo {author} {\bibfnamefont
  {A.~V.}\ \bibnamefont {Andreev}}, \bibinfo {author} {\bibfnamefont
  {E.}~\bibnamefont {Duverger-N\'edellec}},\ and\ \bibinfo {author}
  {\bibfnamefont {L.}~\bibnamefont {Havela}},\ }\bibfield  {title} {\bibinfo
  {title} {Structural, thermodynamic, thermal, and electron transport
  properties of single-crystalline $\mathrm{LaPt}_{2}\mathrm{Si}_{2}$},\ }\href
  {https://doi.org/10.1103/PhysRevB.100.064103} {\bibfield  {journal} {\bibinfo
   {journal} {Phys. Rev. B}\ }\textbf {\bibinfo {volume} {100}},\ \bibinfo
  {pages} {064103} (\bibinfo {year} {2019})}\BibitemShut {NoStop}%
\bibitem [{\citenamefont {Gupta}\ \emph {et~al.}(2017)\citenamefont {Gupta},
  \citenamefont {Dhar}, \citenamefont {Thamizhavel}, \citenamefont {Rajeev},\
  and\ \citenamefont {Hossain}}]{singlecrystalandspecificheat}%
  \BibitemOpen
  \bibfield  {author} {\bibinfo {author} {\bibfnamefont {R.}~\bibnamefont
  {Gupta}}, \bibinfo {author} {\bibfnamefont {S.~K.}\ \bibnamefont {Dhar}},
  \bibinfo {author} {\bibfnamefont {A.}~\bibnamefont {Thamizhavel}}, \bibinfo
  {author} {\bibfnamefont {K.~P.}\ \bibnamefont {Rajeev}},\ and\ \bibinfo
  {author} {\bibfnamefont {Z.}~\bibnamefont {Hossain}},\ }\bibfield  {title}
  {\bibinfo {title} {Superconducting and charge density wave transition in
  single crystalline $\mathrm{LaPt_{2}Si_{2}}$},\ }\href
  {https://doi.org/10.1088/1361-648X/aa70a7} {\bibfield  {journal} {\bibinfo
  {journal} {J. Phys.: Condens. Matter}\ }\textbf {\bibinfo {volume} {29}},\
  \bibinfo {pages} {255601} (\bibinfo {year} {2017})}\BibitemShut {NoStop}%
\bibitem [{\citenamefont {Das}\ \emph {et~al.}(2018)\citenamefont {Das},
  \citenamefont {Gupta}, \citenamefont {Bhattacharyya}, \citenamefont {Biswas},
  \citenamefont {Adroja},\ and\ \citenamefont {Hossain}}]{mutigap-muSR}%
  \BibitemOpen
  \bibfield  {author} {\bibinfo {author} {\bibfnamefont {D.}~\bibnamefont
  {Das}}, \bibinfo {author} {\bibfnamefont {R.}~\bibnamefont {Gupta}}, \bibinfo
  {author} {\bibfnamefont {A.}~\bibnamefont {Bhattacharyya}}, \bibinfo {author}
  {\bibfnamefont {P.~K.}\ \bibnamefont {Biswas}}, \bibinfo {author}
  {\bibfnamefont {D.~T.}\ \bibnamefont {Adroja}},\ and\ \bibinfo {author}
  {\bibfnamefont {Z.}~\bibnamefont {Hossain}},\ }\bibfield  {title} {\bibinfo
  {title} {Multigap superconductivity in the charge density wave superconductor
  $\mathrm{LaPt_{2}Si_{2}}$},\ }\href
  {https://doi.org/10.1103/PhysRevB.97.184509} {\bibfield  {journal} {\bibinfo
  {journal} {Phys. Rev. B}\ }\textbf {\bibinfo {volume} {97}},\ \bibinfo
  {pages} {184509} (\bibinfo {year} {2018})}\BibitemShut {NoStop}%
\bibitem [{\citenamefont {Van~Degrift}(1975)}]{TDOdevice}%
  \BibitemOpen
  \bibfield  {author} {\bibinfo {author} {\bibfnamefont {C.~T.}\ \bibnamefont
  {Van~Degrift}},\ }\bibfield  {title} {\bibinfo {title} {Tunnel diode
  oscillator for 0.001 ppm measurements at low temperatures},\ }\href
  {https://doi.org/10.1063/1.1134272} {\bibfield  {journal} {\bibinfo
  {journal} {Rev. Sci. Instrum.}\ }\textbf {\bibinfo {volume} {46}},\ \bibinfo
  {pages} {599} (\bibinfo {year} {1975})}\BibitemShut {NoStop}%
\bibitem [{\citenamefont {Chia}(2004)}]{TDO2}%
  \BibitemOpen
  \bibfield  {author} {\bibinfo {author} {\bibfnamefont {E.~M.~E.}\
  \bibnamefont {Chia}},\ }\emph {\bibinfo {title} {Penetration depth studies of
  unconventional superconductors}},\ \href
  {http://research.physics.illinois.edu/Publications/theses/copies/Chia_EeMin.pdf}
  {Ph.D. thesis},\ \bibinfo  {school} {University of Illinois at
  Urbana-Champaign,} \bibinfo {year} {2004}\BibitemShut {NoStop}%
\bibitem [{\citenamefont {Prozorov}\ \emph {et~al.}(2000)\citenamefont
  {Prozorov}, \citenamefont {Giannetta}, \citenamefont {Carrington},\ and\
  \citenamefont {Araujo-Moreira}}]{Gfactor}%
  \BibitemOpen
  \bibfield  {author} {\bibinfo {author} {\bibfnamefont {R.}~\bibnamefont
  {Prozorov}}, \bibinfo {author} {\bibfnamefont {R.~W.}\ \bibnamefont
  {Giannetta}}, \bibinfo {author} {\bibfnamefont {A.}~\bibnamefont
  {Carrington}},\ and\ \bibinfo {author} {\bibfnamefont {F.~M.}\ \bibnamefont
  {Araujo-Moreira}},\ }\bibfield  {title} {\bibinfo {title}
  {Meissner-$\mathrm{L}$ondon state in superconductors of rectangular cross
  section in a perpendicular magnetic field},\ }\href
  {https://doi.org/10.1103/PhysRevB.62.115} {\bibfield  {journal} {\bibinfo
  {journal} {Phys. Rev. B}\ }\textbf {\bibinfo {volume} {62}},\ \bibinfo
  {pages} {115} (\bibinfo {year} {2000})}\BibitemShut {NoStop}%
\bibitem [{\citenamefont {Gonnelli}\ \emph {et~al.}(2002)\citenamefont
  {Gonnelli}, \citenamefont {Daghero}, \citenamefont {Ummarino}, \citenamefont
  {Stepanov}, \citenamefont {Jun}, \citenamefont {Kazakov},\ and\ \citenamefont
  {Karpinski}}]{PCS}%
  \BibitemOpen
  \bibfield  {author} {\bibinfo {author} {\bibfnamefont {R.~S.}\ \bibnamefont
  {Gonnelli}}, \bibinfo {author} {\bibfnamefont {D.}~\bibnamefont {Daghero}},
  \bibinfo {author} {\bibfnamefont {G.~A.}\ \bibnamefont {Ummarino}}, \bibinfo
  {author} {\bibfnamefont {V.~A.}\ \bibnamefont {Stepanov}}, \bibinfo {author}
  {\bibfnamefont {J.}~\bibnamefont {Jun}}, \bibinfo {author} {\bibfnamefont
  {S.~M.}\ \bibnamefont {Kazakov}},\ and\ \bibinfo {author} {\bibfnamefont
  {J.}~\bibnamefont {Karpinski}},\ }\bibfield  {title} {\bibinfo {title}
  {Direct $\mathrm{E}$vidence for $\mathrm{T}$wo-$\mathrm{B}$and
  $\mathrm{S}$uperconductivity in $\mathrm{Mg}\mathrm{B}_{2}$ $\mathrm{S}$ingle
  $\mathrm{C}$rystals from $\mathrm{D}$irectional
  $\mathrm{P}$oint-$\mathrm{C}$ontact $\mathrm{S}$pectroscopy in
  $\mathrm{M}$agnetic $\mathrm{F}$ields},\ }\href
  {https://doi.org/10.1103/PhysRevLett.89.247004} {\bibfield  {journal}
  {\bibinfo  {journal} {Phys. Rev. Lett.}\ }\textbf {\bibinfo {volume} {89}},\
  \bibinfo {pages} {247004} (\bibinfo {year} {2002})}\BibitemShut {NoStop}%
\bibitem [{\citenamefont {Daghero}\ and\ \citenamefont
  {Gonnelli}(2010)}]{PCS2}%
  \BibitemOpen
  \bibfield  {author} {\bibinfo {author} {\bibfnamefont {D.}~\bibnamefont
  {Daghero}}\ and\ \bibinfo {author} {\bibfnamefont {R.~S.}\ \bibnamefont
  {Gonnelli}},\ }\bibfield  {title} {\bibinfo {title} {Probing multiband
  superconductivity by point-contact spectroscopy},\ }\href
  {https://doi.org/10.1088/0953-2048/23/4/043001} {\bibfield  {journal}
  {\bibinfo  {journal} {Supercond. Sci. Technol.}\ }\textbf {\bibinfo {volume}
  {23}},\ \bibinfo {pages} {043001} (\bibinfo {year} {2010})}\BibitemShut
  {NoStop}%
\bibitem [{\citenamefont {Gupta}\ \emph {et~al.}(2018)\citenamefont {Gupta},
  \citenamefont {Rajeev},\ and\ \citenamefont {Hossain}}]{thermaltransport}%
  \BibitemOpen
  \bibfield  {author} {\bibinfo {author} {\bibfnamefont {R.}~\bibnamefont
  {Gupta}}, \bibinfo {author} {\bibfnamefont {K.~P.}\ \bibnamefont {Rajeev}},\
  and\ \bibinfo {author} {\bibfnamefont {Z.}~\bibnamefont {Hossain}},\
  }\bibfield  {title} {\bibinfo {title} {Thermal transport studies on charge
  density wave materials $\mathrm{LaPt_{2}Si_{2}}$ and
  $\mathrm{PrPt_{2}Si_{2}}$},\ }\href
  {http://stacks.iop.org/0953-8984/30/i=47/a=475603} {\bibfield  {journal}
  {\bibinfo  {journal} {J. Phys.: Condens. Matter}\ }\textbf {\bibinfo {volume}
  {30}},\ \bibinfo {pages} {475603} (\bibinfo {year} {2018})}\BibitemShut
  {NoStop}%
\bibitem [{\citenamefont {Carrington}\ and\ \citenamefont
  {Manzano}(2003)}]{delta0}%
  \BibitemOpen
  \bibfield  {author} {\bibinfo {author} {\bibfnamefont {A.}~\bibnamefont
  {Carrington}}\ and\ \bibinfo {author} {\bibfnamefont {F.}~\bibnamefont
  {Manzano}},\ }\bibfield  {title} {\bibinfo {title} {Magnetic penetration
  depth of $\mathrm{MgB}_2$},\ }\href
  {https://doi.org/https://doi.org/10.1016/S0921-4534(02)02319-5} {\bibfield
  {journal} {\bibinfo  {journal} {Phys. C (Amsterdam, Neth.)}\ }\textbf {\bibinfo {volume}
  {385}},\ \bibinfo {pages} {205 } (\bibinfo {year} {2003})}\BibitemShut
  {NoStop}%
\bibitem [{\citenamefont {Prozorov}\ and\ \citenamefont
  {Giannetta}(2006)}]{gk}%
  \BibitemOpen
  \bibfield  {author} {\bibinfo {author} {\bibfnamefont {R.}~\bibnamefont
  {Prozorov}}\ and\ \bibinfo {author} {\bibfnamefont {R.~W.}\ \bibnamefont
  {Giannetta}},\ }\bibfield  {title} {\bibinfo {title} {Magnetic penetration
  depth in unconventional superconductors},\ }\href
  {https://doi.org/10.1088/0953-2048/19/8/r01} {\bibfield  {journal} {\bibinfo
  {journal} {Supercond. Sci. Technol.}\ }\textbf {\bibinfo {volume} {19}},\
  \bibinfo {pages} {R41} (\bibinfo {year} {2006})}\BibitemShut {NoStop}%
\bibitem [{\citenamefont {Blonder}\ \emph {et~al.}(1982)\citenamefont
  {Blonder}, \citenamefont {Tinkham},\ and\ \citenamefont {Klapwijk}}]{BTK}%
  \BibitemOpen
  \bibfield  {author} {\bibinfo {author} {\bibfnamefont {G.~E.}\ \bibnamefont
  {Blonder}}, \bibinfo {author} {\bibfnamefont {M.}~\bibnamefont {Tinkham}},\
  and\ \bibinfo {author} {\bibfnamefont {T.~M.}\ \bibnamefont {Klapwijk}},\
  }\bibfield  {title} {\bibinfo {title} {Transition from metallic to tunneling
  regimes in superconducting microconstrictions: Excess current, charge
  imbalance, and supercurrent conversion},\ }\href
  {https://doi.org/10.1103/PhysRevB.25.4515} {\bibfield  {journal} {\bibinfo
  {journal} {Phys. Rev. B}\ }\textbf {\bibinfo {volume} {25}},\ \bibinfo
  {pages} {4515} (\bibinfo {year} {1982})}\BibitemShut {NoStop}%
\bibitem [{\citenamefont {Sasaki}\ \emph {et~al.}(2011)\citenamefont {Sasaki},
  \citenamefont {Kriener}, \citenamefont {Segawa}, \citenamefont {Yada},
  \citenamefont {Tanaka}, \citenamefont {Sato},\ and\ \citenamefont
  {Ando}}]{CuBiSe}%
  \BibitemOpen
  \bibfield  {author} {\bibinfo {author} {\bibfnamefont {S.}~\bibnamefont
  {Sasaki}}, \bibinfo {author} {\bibfnamefont {M.}~\bibnamefont {Kriener}},
  \bibinfo {author} {\bibfnamefont {K.}~\bibnamefont {Segawa}}, \bibinfo
  {author} {\bibfnamefont {K.}~\bibnamefont {Yada}}, \bibinfo {author}
  {\bibfnamefont {Y.}~\bibnamefont {Tanaka}}, \bibinfo {author} {\bibfnamefont
  {M.}~\bibnamefont {Sato}},\ and\ \bibinfo {author} {\bibfnamefont
  {Y.}~\bibnamefont {Ando}},\ }\bibfield  {title} {\bibinfo {title}
  {Topological superconductivity in
  $\mathrm{Cu}_{x}\mathrm{Bi}_{2}\mathrm{Se}_{3}$},\ }\href
  {https://doi.org/10.1103/PhysRevLett.107.217001} {\bibfield  {journal}
  {\bibinfo  {journal} {Phys. Rev. Lett.}\ }\textbf {\bibinfo {volume} {107}},\
  \bibinfo {pages} {217001} (\bibinfo {year} {2011})}\BibitemShut {NoStop}%
\bibitem [{\citenamefont {Sheet}\ \emph {et~al.}(2004)\citenamefont {Sheet},
  \citenamefont {Mukhopadhyay},\ and\ \citenamefont
  {Raychaudhuri}}]{criticalcurrent}%
  \BibitemOpen
  \bibfield  {author} {\bibinfo {author} {\bibfnamefont {G.}~\bibnamefont
  {Sheet}}, \bibinfo {author} {\bibfnamefont {S.}~\bibnamefont
  {Mukhopadhyay}},\ and\ \bibinfo {author} {\bibfnamefont {P.}~\bibnamefont
  {Raychaudhuri}},\ }\bibfield  {title} {\bibinfo {title} {Role of critical
  current on the point-contact Andreev reflection spectra between a normal
  metal and a superconductor},\ }\href
  {https://doi.org/10.1103/PhysRevB.69.134507} {\bibfield  {journal} {\bibinfo
  {journal} {Phys. Rev. B}\ }\textbf {\bibinfo {volume} {69}},\ \bibinfo
  {pages} {134507} (\bibinfo {year} {2004})}\BibitemShut {NoStop}%
\bibitem [{\citenamefont {Brandt}(1976)}]{MagneticFieldFormula}%
  \BibitemOpen
  \bibfield  {author} {\bibinfo {author} {\bibfnamefont {E.~H.}\ \bibnamefont
  {Brandt}},\ }\bibfield  {title} {\bibinfo {title} {Microscopic theory of
  clean type-\uppercase\expandafter{\romannumeral2} superconductors in the
  entire field-temperature plane},\ }\href
  {https://doi.org/10.1002/pssb.2220770109} {\bibfield  {journal} {\bibinfo
  {journal} {Phys. Status Solidi B}\ }\textbf {\bibinfo {volume} {77}},\
  \bibinfo {pages} {105} (\bibinfo {year} {1976})}\BibitemShut {NoStop}%
\end{thebibliography}
\end{document}